\documentclass[preprint,10pt]{aastex}
\usepackage{lscape}

\accepted{}
\journalid{}{}
\articleid{}{}

\shortauthors{Mohanty et al.}
\shorttitle{Environment and Temperature of Young Substellar EB}
\slugcomment{To appear in ApJ}

\begin{document}

\def\lam{$\lambda$}
\def\tross{${\tau}_{R}$ }
\def\hal{{{\rm H}\alpha} }
\def\rc{R_C}
\def\ic{I_C}
\def\rcm{R_{Cm}}
\def\icm{I_{Cm}}
\def\jm{J_{m}}
\def\rco{R_{Co}}
\def\ico{I_{Co}}
\def\jo{J_{o}}
\def\av{A_V}
\def\ar{A_{R_C}}
\def\ai{A_{I_C}}
\def\exj{A_{J}}
\def\kr{k_{R_C}}
\def\ki{k_{I_C}}
\def\rad{{\mathcal{R}}_{\ast}}
\def\dist{\mathcal{D}}
\def\mass{{\mathcal{M}}_{\ast}}
\def\lum{{\mathcal{L}}_{bol}}
\def\logten{log_{10}}

\def\li{\ion{Li}{1}\ }
\def\na{\ion{Na}{1}\ }
\def\pot{\ion{K}{1}\ }
\def\oxy{\ion{O}{1}\ }
\def\hel{\ion{He}{1}\ }
\def\cal{\ion{Ca}{2}\ }
\def\nit{\ion{N}{2}\ }
\def\sul{\ion{S}{2}\ }

\def\ualph{${\mu}_{\alpha}$ }
\def\udelt{${\mu}_{\delta}$ }
\def\kms{km s$^{-1}$\ }
\def\kmsp{km s$^{-1}$ pix$^{-1}$\ }
\def\cms{cm s$^{-1}$\ }
\def\cmc{cm$^{-3}$\ }
\def\cmss{cm$^{2}$ s$^{-1}$\ }
\def\cmcs{cm$^{3}$ s$^{-1}$\ }

\def\mdot{$\dot{M}$ }
\def\msun{M$_\odot$ }
\def\rsun{R$_\odot$ }
\def\lsun{L$_\odot$ }
\def\mj{M$_{Jup}$ }

\def\teff{$T_{\rm eff}$}
\def\tefflbol{T$_{e\! f\! f}$-L$_{\it bol}$~}
\def\gv{{\it g}~}
\def\vsini{{\it v}~sin{\it i}~}
\def\vrad{v$_{\it rad}$~}
\def\lbol{L$_{\it bol}$~}
\def\mbol{{\rm M}_{\it bol}}
\def\lhal{L_{H\alpha}}
\def\fhal{F_{H\alpha}}
\def\fcal{{\mathcal{F}}_{CaII}}
\def\fcont{{\mathcal{F}}_{cont}}
\def\fbol{F_{\it bol}}
\def\lx{L_{X} }
\def\eqwhal{EW_{H\alpha}}
\def\eqwcal{EW_{CaII}}
\def\alom{{\alpha}{\Omega} }
\def\ross{R_{0} }
\def\cots{{\tau}_{c} }
\def\fchal{{\mathcal{F}}_{c\hal} }
\def\h2o{H$_2$O}
\def\vcal{r_{CaII} }
\def\fex{{\mathcal{F}}_{excess}}
\def\eb{2M0535$-$05}

\title{Circumstellar Environment and Effective Temperature of
the Young Substellar Eclipsing Binary 2MASS~J05352184$-$0546085}

\author{Subhanjoy Mohanty\altaffilmark{1}, Keivan G.\ Stassun\altaffilmark{2},
Robert D.\ Mathieu\altaffilmark{3}}

\altaffiltext{1}{Harvard-Smithsonian Center for Astrophysics, Cambridge, MA 02138, USA.  smohanty@cfa.harvard.edu}
\altaffiltext{2}{Department of Physics \& Astronomy, Vanderbilt University, Nashville, TN 37235, USA.}
\altaffiltext{3}{Department of Astronomy, University of Wisconsin--Madison, Madison, WI 53706, USA.}

\begin{abstract} 
We present new {\it Spitzer} IRAC/PU/MIPS photometry from 3.6 to 24
$\mu$m, and new Gemini GMOS photometry at 0.48 $\mu$m, of the young
brown dwarf eclipsing binary 2MASS~J05352184$-$0546085, located in the
Orion Nebula Cluster.  No excess disk emission is detected: The measured
fluxes at $\lambda \le 8 \mu {\rm m}$ are within $1\sigma$ 
($\lesssim 0.1$ mJy)
of a bare photosphere, and the $3\sigma$ upper limit at 16 $\mu$m is 
a mere 0.04 mJy above the bare photospheric level.  Together
with the known properties of the system, this implies the absence of
optically thick disks around the individual components.  It also
implies that if any circumbinary disk is present, it must either be optically thin and extremely tenuous (10$^{-10}$ \msun) if it extends in to within $\sim$0.1 AU of the binary (the approximate tidal truncation radius), or it must be optically thick with a large inner hole, $>$0.6--10 AU in radius depending on degree of flaring. The consequence in all cases is that disk accretion is likely to be negligible or absent.  This supports the recent proposal that the strong $\hal$ emission in
the primary (more massive) brown dwarf results from chromospheric
activity, and thereby bolsters the hypothesis that the surprising
\teff\ inversion observed between the components is due to strong
magnetic fields on the primary.  Our data also set constraints on the
\teff\ of the components independent of spectral type, and thereby
on models of the aforementioned magnetic field effects.  We discuss the
consequences for the derived fundamental properties of young brown
dwarfs and very low-mass stars in general.  
Specifically, if very active isolated young brown dwarfs and very
low-mass stars suffer the same activity/field related effects as the
\eb\ primary, the low-mass stellar/substellar IMF currently derived
from standard evolutionary tracks may be substantially in error.
\end{abstract}


\section{Introduction}

The very young system \objectname{2MASS~J05352184$-$0546085}
(henceforth \eb), located in the Orion star-forming region, has been
identified by \citet[][hereafter SMV06]{stassun06} as the first known
substellar eclipsing binary (EB). EBs allow exquisitely precise
direct determinations of the component masses and radii, as well
the ratio of their surface brightnesses (or equivalently, ratio of
their \teff). As such, \eb\ allows the first stringent tests of the
theoretical evolutionary models widely employed to characterize the
vast majority of brown dwarfs (for which direct measurements of mass
and radius are not possible).

The analysis of SMV06 \citep[and the follow-up analysis
of][]{stassun07} reveals that: {\it (1)} both components of \eb\
are moderate-mass brown dwarfs ($0.057 \pm 0.004$ and 
$0.036 \pm 0.003$ M$_{\odot}$); {\it
(2)} their radii are $\sim$10\% larger than expected from standard
evolutionary models for their assumed age of 1 Myr \citep[see Fig.~3
of][]{chabrier07}; and {\it (3)} the ratio of the secondary's to
primary's \teff\ is $T_2/T_1 \approx 1.05$, i.e., the more massive
primary is {\it cooler} than its less massive companion.  This last
point is highly surprising, since all standard models for substellar
evolution predict that for coeval bodies (as the EB components are
likely to be; but see \S5.2), higher mass translates to higher \teff,
not lower.

The most compelling explanation for the \teff\ reversal (and
radius discrepancy) has been advanced by \citet[][henceforth
CGB07]{chabrier07}, who propose that it arises from the suppression of
convection by rapid rotation and strong magnetic fields. In particular,
they show that a significant reduction in convective efficiency
combined with a large areal coverage of cool magnetic spots, in the
primary, can reproduce SMV06's results; as a by-product, they predict
\teff s of 2320K and 2440K for the primary and secondary, respectively.
This hypothesis has been bolstered via optical spectroscopy by
\citet[][henceforth R07]{reiners07}, who find that the primary is
indeed rapidly rotating and evinces strong $\hal$ emission indicative
of large surface magnetic fields (\vsini $\sim$ 10 \kms; $\hal$
equivalent width (EW) $\gtrsim$ 32.6\AA\ $\Rightarrow$ $\lhal$/\lbol
$\gtrsim -3.47$ $\Rightarrow$ magnetic field strength $Bf$ $\sim$
4 kG), while the secondary is a slow rotator with comparatively weak
$\hal$ emission/magnetic fields (\vsini $<$ 5 \kms; $\hal$ EW $\sim$
4.8\AA\ $\Rightarrow$ $\lhal$/\lbol $\sim -4.30$ $\Rightarrow$  $Bf$
$\sim$ 2 kG).

Nevertheless, two open questions remain.  First, R07's support
of CGB07's theory depends on ascribing the strong $\hal$ emission
in the primary to chromospheric activity \citep[which allows R07
to translate it to a magnetic field strength using the field dwarf
$\lhal$/\lbol--$B$ relationship found by][]{reinersbasri07}.  However,
\eb\ is an extremely young system in a star-forming region; like stars,
brown dwarfs at such ages very often exhibit strong $\hal$ emission due
to {\it disk accretion} instead of activity \citep[e.g.][]{Mohanty05,
muzerolle05}.  The contamination of the primary's instrinsic emission
by the strong nebular emission (which permits R07 to set only a lower
limit on its $\hal$ EW) means that the line profile and EW cannot
categorically exclude accretion; another diagnostic is required.

Second, the absolute \teff s of the components of \eb\ have yet
to be accurately determined. SMV06's eclipse data do not allow a
direct measure of the individual \teff\ of the components, only their
ratio, so spectral type information was used to estimate the separate
temperatures. From optical and near-infrared (NIR) spectra and colors,
Stassun et al. (2007, henceforth S07) find spectral types of 
$\sim$ M6.5--M7 for
the components, and thereby \teff\ $\sim$ 2700K and $\sim$ 2900K for
the primary and secondary respectively.  R07 further find these temperatures to be consistent with detailed analysis of TiO bands in high-resolution optical spectra.  CGB07, on the other hand, predict \teff s significantly cooler, by $\sim$300K, for both (see above). Temperatures derived independently of
spectral types can thus validate/constrain both the CGB07 models
and the spectral type--\teff\ conversion scale in the presence of
suppressed convection.

Detailed spectral energy distribution (SED) observations and modeling
are ideal for investigating both questions.  For example, a lack of
significant mid-infrared (MIR) excess dust emission would strongly
argue against any surrounding optically thick accretion disks in \eb,
and thus against ongoing accretion, making the case for chromospheric
activity in the primary more robust. Simultaneously, detection of
photospheric levels from the optical to the MIR (in the absence of
disk emission) would allow constraints on the \teff s via more general
blackbody/opacity considerations than spectral type arguments.

To accomplish these goals, in this paper we combine NIR photometry from
the literature with new {\it Spitzer} photometry over 3.6--24 $\mu$m and
Gemini photometry at 0.48 $\mu$m, to construct a complete SED of \eb\ from
0.48 $\mu$m to 24 $\mu$m (\S2).  In \S3 we present synthetic SED models
of substellar objects surrounded by disks with various geometries, for the
purpose of comparing against the observed SED of \eb. Results are presented
in \S4, where we find that the observations clearly indicate that \eb\
is devoid of optically thick disk dust within at least $\sim$ 0.6--10AU of the
brown dwarfs, sugesting that accretion is negligible or absent in \eb. We
also find that the observed SED is most consistent with a relatively warm
temperature scale for the brown dwarfs, similar to that found for field
M dwarfs. As we discuss in \S5, the observational evidence now strongly
implicates magnetic activity on the primary (more massive) brown dwarf as
the reason for the surprising reversal of temperatures in \eb. However,
the CGB07 models of young brown dwarfs undergoing magnetically suppressed
convection predict \teff s several hundred K cooler than inferred
from the observed SED, and in \S5 we discuss possible ways to rectify
this discrepancy.

\section{Observations}

We constructed the spectral energy distribution (SED) of \eb\
from broadband flux measurements over the wavelength range 0.48--24
$\mu$m. These measurements are summarized in Table 1. In all cases
we confirmed that the observations were made outside eclipse, to
ensure that the full combined flux of both components was obtained. In
addition, as noted below, we adopt uncertainties somewhat larger than
the formal measurement errors to account for systematic errors and/or
spot-induced variability in \eb.

\subsection{Photometry from the literature}

NIR magnitudes at $JHK_S$ (1.2--2.2 $\mu$m) were taken
from the 2MASS database and converted to fluxes using the
zero-points described in the 2MASS All-Sky Data Release Explanatory
Supplement\footnote{\url{http://www.ipac.caltech.edu/2mass/releases/allsky/doc/sec6\_4a.html}}.
The formal uncertainties in each passband are $\sim 0.02$ mag. We adopt
uncertainties of 0.05 mag to account for out-of-eclipse variability
at these wavelengths \citep{gomez08}. An optical flux at $I_C$ was
taken from the observations of \citet{stassun99}, where we adopt
an uncertainty of 0.1 mag to account for out-of-eclipse variability
observed in the $I$ band (S07). 

\subsection{New photometry}
A flux measurement in the Sloan $g'$ passband was obtained with the
Gemini North GMOS instrument under photometric conditions on 2008 Jan
24 with an exposure time of 300~s. There are few suitable comparison
stars in the small GMOS field of view. We selected the star Par~1812
as the comparison star because it is reasonably bright but unsaturated
in our observations, and has both $B$ and $V$ magnitudes available
in the SIMBAD catalog. \eb\ was measured to be 5.93 mag fainter than
Par~1812 in the $g'$ filter in this observation, and its $g'$ magnitude
converted to absolute flux units using the zero-points defined in
\citet{fukugita96}. The formal measurement error is $\sim 0.02$ mag. We
adopt an uncertainty of 0.1 mag to account for intrinsic variability
in the comparison star of $\sim 0.1$ mag in the visible \citep{gcvs}.

To cover the MIR portion of the SED where any disk emission would be
expected to dominate, new imaging observations of \eb\ were obtained
with the {\it Spitzer} Space Telescope (program \#40503). We observed
at all four passbands of the IRAC instrument (3.6--8 $\mu$m), with
the IRS instrument in blue peak-up mode (16 $\mu$m), and with the
MIPS instrument at 24 $\mu$m. Forty IRAC frames were obtained in
High Dynamic Range mode in a medium-scale dither pattern. One 12-s
long exposure and one 0.6-s short exposure were obtained for each
frame in all 4 channels, yielding total exposure times of 480~s and
24~s for the long and short exposures respectively. This allowed
us to avoid overexposing in the more sensitive channels 1 and 2
(3.6 and 4.5 $\mu$m) while obtaining sufficient S/N in the less
sensitive channels 3 and 4 (5.8 and 8.0 $\mu$m). Fifty dithered IRS
blue peak-up images were obtained with a 30~s integration time for
each, for a total on-source integration time of 1500~s. Finally, 24
$\mu$m MIPS photometry was acquired in the small field of view mode,
with 20 cycles, 15 frames per cycle and 10-s frames, resulting in a
total on-source integration time of 3000~s.

\eb\ was detected in all four of the {\it Spitzer} IRAC
images. We followed the procedure described in the IRAC Data
Handbook\footnote{\url{http://ssc.spitzer.caltech.edu/irac/dh/}}
for performing aperture photometry on the pipeline generated mosaic
images. Specifically, we used an aperture radius of 3 pixels and
a background annulus of 3--7 pixels, and then applied aperture
correction factors of 1.124, 1.127, 1.143, and 1.234 in IRAC channels
1, 2, 3, and 4, respectively. Photometry for channels 1 and 2 was
performed on the short exposure images, and for channels 3 and 4 on
the long exposures. The measurement uncertainties of 10--50\% (Table
\ref{obs-table}) are dominated by the bright and highly spatially
variable nebular background.

\eb\ was not detected in the IRS peak-up (16 $\mu$m) and MIPS 24
$\mu$m images. Thus we instead determined upper limits by measuring
the per-pixel background noise level in an annulus centered on
the source position, with inner and outer radii of 1 and 5 pixels,
respectively. These too are reported in Table~\ref{obs-table}.

The resulting observed SED of \eb\ at 0.48--24 $\mu$m is displayed in
Fig.~\ref{obs-sed-fig}. For comparison, the figure also shows a model SED
using the {\sc dusty} model atmospheres of \citet{Allard01} at the radii
and temperatures determined by S07 ($R_1=0.675\; {\rm R}_\odot$, 
$R_2=0.486\; {\rm R}_\odot$, $T_1=2715$~K, $T_2=2890$~K). The bolometric 
luminosities are computed directly from the temperatures and radii (via 
the Stefan-Boltzmann relation). The model SED was fit to the data via
$\chi^2$ minimization, where the only free parameters are the distance and
the extinction, for which we find $d = 415 \pm 19$ pc and $A_V = 0.4 \pm
0.2$ mag. The reduced $\chi^2$ of the fit is $\chi_\nu^2 = 0.97$ 
($\nu = 7$ degrees of freedom). 
These distance and extinction values are consistent with other 
determinations to the young Orion Nebula Cluster 
\citep[ONC; e.g.,][]{hirota07,genzel81,hillenbrand97}.

\section{Models Employed: Disks and Synthetic Spectra}

To compare the observed SED of \eb\ with that expected from young
brown dwarfs with disks, we employed the Monte Carlo radiative
transfer code of \citet{whitney03a,whitney03b} to generate model
SEDs. The code randomly emits photons from the central illuminating
source(s) and follows the photons as they interact with (i.e., are
absorbed or scattered by) any circumstellar material. Here we model
the circumstellar material as an optically thick circumbinary disk of
dust extending from an inner truncation radius, $R_{\rm trunc}$, to an
outer radius of 100 AU. The code self-consistently solves for thermal
equilibrium in the disk as absorbed photons heat the disk and are
re-radiated. Sublimation of dust is also included \citep[for details
of the dust properties used by the code, see Table 3 of][]{whitney03b}.

The most important input parameters of the central illuminating
source(s) are the total luminosity and the intrinsic
(photospheric) SED. We adopt the effective temperatures and
luminosities for the components of \eb\ from S07. Built into 
the code are the solar-metallicity {\sc NextGen} atmosphere models of
\citet{hauschildt99}, which are virtually identical to the {\sc dusty}
models at the warm \teff\ of S07.

We considered two classes of disk models, corresponding to different
flaring assumptions:
\begin{enumerate}
\item Flared disk, where disk scale-height $h$ increases 
with radial distance $\varpi$ from the star: $h = h_0
\varpi^{1.25}$ ($h_0$ = scale-height at stellar surface determined
via hydrostatic equilibrium calculation, typically $\sim
0.01 R_\star$).  This corresponds to a disk in vertical hydrostatic
equilibrium and well-mixed gas and dust, with flaring caused
by heating of grains in the optically thin disk surface by stellar
irradiation.


\item Geometrically thin, flat disk, where $h$ remains 
constant with $\varpi$.  This
corresponds to the case where all the grains
have settled to the disk mid-plane. 
\end{enumerate}

\noindent In all models (except one case in \S 4.1 where explicitly stated otherwise) we used a disk mass of $10^{-3}\;
{\rm M}_\odot$ with a surface density profile falling off like
$\varpi^{-2.25}$ \citep[see][]{whitney03a,whitney03b}. In addition,
the disks are modeled as ``passive" disks, i.e.\ they do not generate
any intrinsic lumonisity via accretion. As such, the fluxes predicted
by these model SEDs should be regarded as the {\it minimum} fluxes that
such disks would be expected to produce. The resulting model SEDs for
the two disk classes above are shown in Figs.~\ref{knife-sed-fig}
and \ref{flared-sed-fig}.

\section{Results}
\subsection{No evidence for accretion disks in \eb}

Figs.~\ref{knife-sed-fig}--\ref{flared-sed-fig} show that, if the disk is optically thick and viewed edge-on (i.e., in the orbital plane of the eclipsing binary, as expected from the standard assumption of co-planarity), then only a
geometrically thin flat disk can fit the
observed flux measurements at 0.48--8.0$\mu$m and upper limits at
16--24$\mu$m \footnote{Strictly speaking, flat disk models consistent with the data can only
fit the 16$\mu$m upper limit; while remaining consistent with the 24$\mu$m
upper limit, they imply that the true 24$\mu$m flux must be much lower.
Any flat disk model that passes through the 24$\mu$m upper limit produces a 16$\mu$m flux much higher than the observed upper limit in the latter band, and can thus be ruled out.}. Moreover, such a flat disk,
if present, must also possess a large inner hole at least 0.6 AU in
radius.  

Dropping the coplanarity assumption yields even larger inner holes (not shown): as the disk is seen more face-on, the observed flux from it increases, requiring a larger hole to bring the flux back down to the observed values / upper limits. 
    


An optically thick ``flared'' disk can only fit the data if it is severely non-coplanar with the binary orbital plane {\it and} has a very large inner hole (Fig.\ref{flared-sed-fig}). If seen edge-on, the relatively large vertical scale-heights of the outer regions of such a disk produce $\gtrsim 5$ mag of extinction at 1 $\mu$m, and even
more at shorter wavelengths. (The model SEDs show strong sampling noise
in Fig.~\ref{flared-sed-fig} due to the severe
attenuation of photons from the brown dwarfs at these wavelengths.) 
Thus to match the observed optical and NIR SED, it
needs to be seen at a much more face-on orientation. As in the flat disk case, however, the observed MIR flux increases with such orientation (and more severely so for the flared disk than the flat, since the former intercepts more stellar radiation than the latter for a given hole size).  This again pushes the hole inner radius to even larger values to remain consistent with the MIR data.  The result, as shown in Fig.~\ref{flared-sed-fig}, is that an optically thick flared disk can only fit the observed SED of \eb\ if it is inclined to the orbital plane by $45^\circ$ and simultaneously possesses an inner hole at least $\sim 10$ AU in radius. 

All the above immediately also implies that no optically thick disk exists around the individual components, since the smallest hole size the data can accomodate for such a disk is 0.6 AU, while the binary semi-major axis is only 0.04 AU (S07).  Thus an optically thick disk, if at all present, must be circumbinary.

The results also imply that any disk material that extends inwards of 0.6 AU must be optically thin.  We explore this case further by assuming an inner truncation radius equal to the tidal truncation radius for the binary.  Based on the simulations by \citet{al96}, we take this to be $\sim$3 times the semi-major axis (0.04 AU) $\approx$ 0.1 AU.  For a flat disk with a hole of this size, and the same disk outer radius and radial density profile as before, we find the maximum dust-disk mass consistent with the data to be 10$^{-10}$ \msun (i.e., a paltry 10$^{-9}$ of the total binary mass).  Note that a smaller inner hole would imply an even smaller disk mass; as a corollary, any optically thin disks around the individual components must also be less massive than this limit.       


Accretion can be ruled out for the optically thin disk cases: the extremely low dust masses implied for these, $<$ 3$\times$10$^{-3}$ lunar masses, are far too small for primordial accretion disks.  For the optically thick cases, the lower limits on inner hole size of 0.6-10 AU also makes accretion unlikely.  Even the smaller value of 0.6 AU equals $\gtrsim$6 times the tidal truncation radius and $\sim$30 times the dust destruction radius.  
Such a large hole in the dust disk may be cleared either by substantial
grain growth, or by sweeping of material by planet/planetesimal formation,
or by photoevaporation.  The first would imply very significant grain
evolution, suggesting that the disk has probably moved beyond the main
accretion phase.  The other two scenarios would result in the clearing of
gas as well as dust within the hole.  All three scenarios therefore make
significant ongoing accretion unlikely.

Moreover, if accretion streams were indeed spiralling in from the inner edge of the circumbinary disk onto the binary, the theory and simulations of \citet{al96} indicate that they should preferentially land on the less massive component, manifestly {\it not} the situation in \eb\ if the observed $\gtrsim 7\times$ stronger $\hal$ emission in the primary \citep{reiners07} is taken as an accretion
indicator. Additionally, \eb\ exhibits no large-amplitude photometric
variability (S07), in contrast to that seen in the young
binaries so far shown to be experiencing accretion from circumbinary
disks \citep{mathieu97,jensen07}.

Finally, the data do not explicitly {\it require} the presence of
a disk at all. The data out to 8 $\mu$m in fact indicate a naked photosphere alone, and it is only the uncertainty embodied in the 16 and 24$\mu$m upper
limits (and the lack of data beyond) that forces a conservative
inclusion of the possibility of a disk.  All of the evidence taken
together therefore suggests that significant ongoing gas accretion in this
system is very unlikely, and that the strong $\hal$ emission observed
in the primary is indeed related to chromospheric activity as suggested
by R07.

\subsection{Effective temperatures of \eb\ are higher than predicted
by CGB07 models}
We now discuss the \teff\ of the 2M0535-05 components in more detail,
by comparing the observed SED to {\sc dusty} synthetic spectra.
We note that for the hotter \teff\ proposed by S07, {\sc nextgen}
spectra (which {\it a priori} neglect dust formation) are very similar
to {\sc dusty} ones, since no dust forms at these relatively warm
\teff\ anyway.  However, the cooler ones proposed by CGB07 would lead
to dust formation.  For consistency, therefore, we use {\sc dusty}
models for evaluating both the the S07 and CGB07 temperatures.

\begin{figure}[ht]
\epsscale{0.8}
\plotone{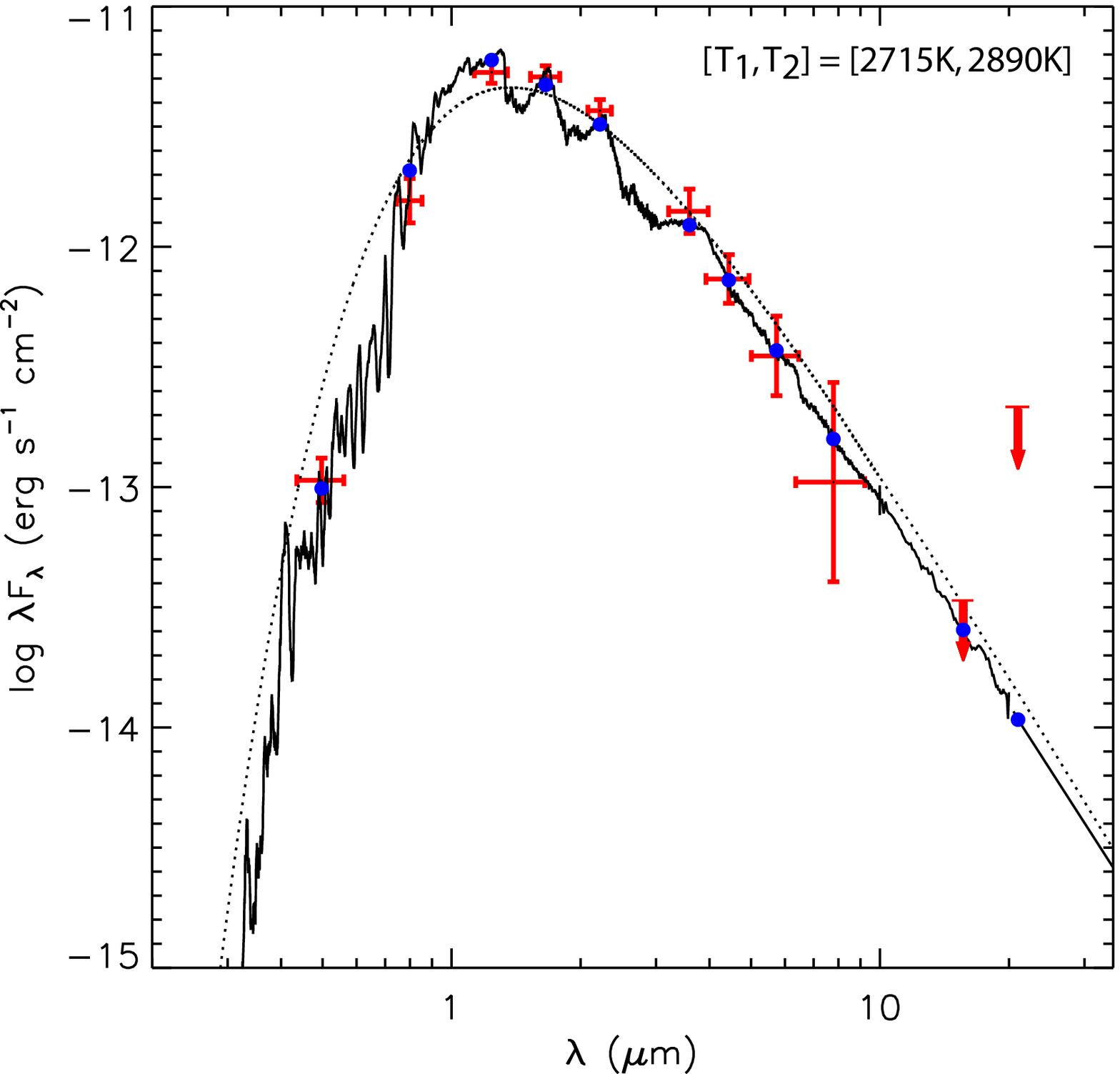}
\caption{\label{obs-sed-fig}
Observed spectral energy distribution (SED) of \eb\ from 0.48
$\mu$m to 24 $\mu$m. Measured fluxes (Table \ref{obs-table})
are shown with $1\sigma$ vertical error bars and horizontal bars
representing the filter passbands. Downward arrows represent $3\sigma$
upper limits (the upper limit is the horizontal bar on the tail of the arrow). 
Circles represent the predicted model fluxes for each 
of the observed bandpasses. The solid curve is a model SED constructed 
from the {\sc dusty} model atmospheres of \citet{Allard01}, with component
temperatures of 2715~K and 2890~K for the primary (more massive) and
secondary brown dwarf, respectively (S07; R07). The model SED
has been scaled and reddened by best-fit values of 415 pc and $A_V =
0.4$ mag. The dotted curve represents the SED of pure blackbodies
corresponding to the above temperatures.
}
\end{figure}

\begin{figure}[ht]
\epsscale{0.8}
\plotone{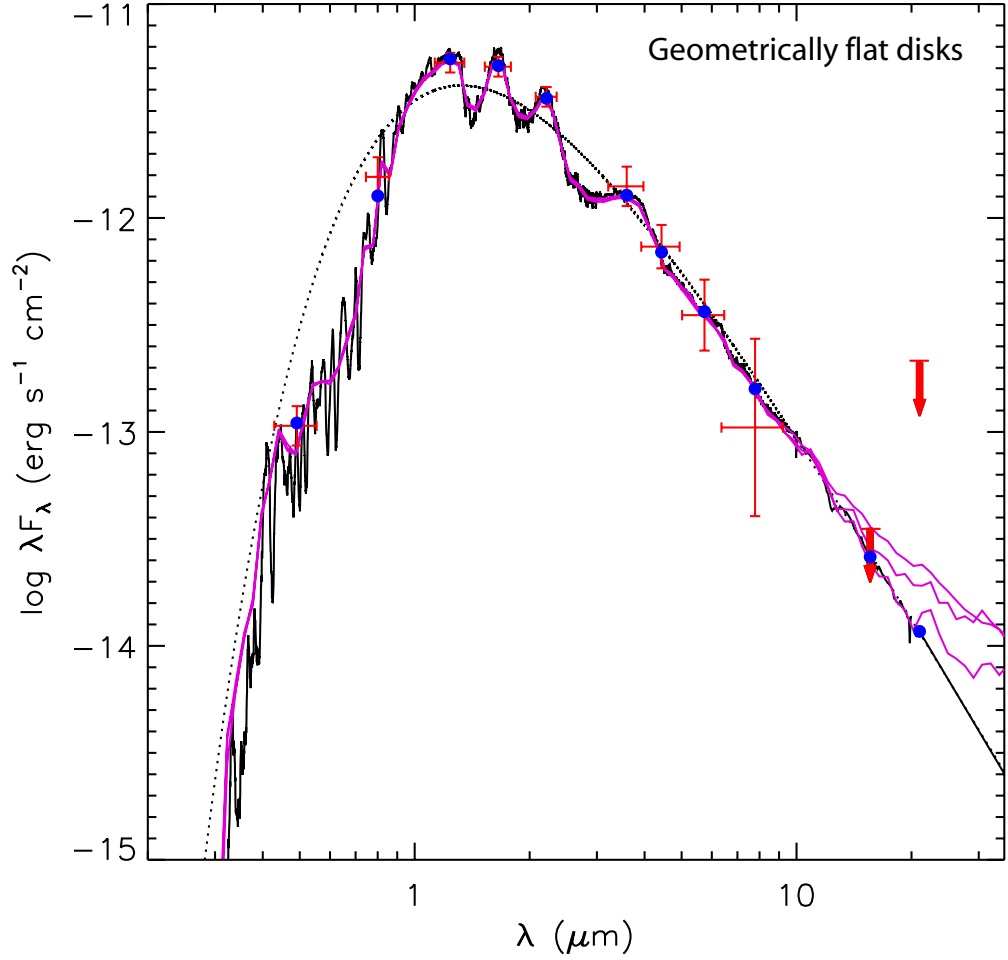}
\caption{\label{knife-sed-fig}
Model SEDs of geometrically flat disks with inner truncation radii of 0.6, 
1, and 10 AU are shown as magenta solid curves; smaller inner radius corresponds to larger mid-IR flux.  The 0.6 AU hole is the smallest possible that is still consistent with the observed upper limit at 16 $\mu$m.  
The models are shown at an inclination of $88^\circ$ relative to the line of sight, i.e.\ nearly edge-on, corresponding to the measured orbital inclination of the central eclipsing binary (S07). Symbols and black curves are
as in Fig.~\ref{obs-sed-fig}. 
}
\end{figure}

\begin{figure}[ht]
\epsscale{0.8}
\plotone{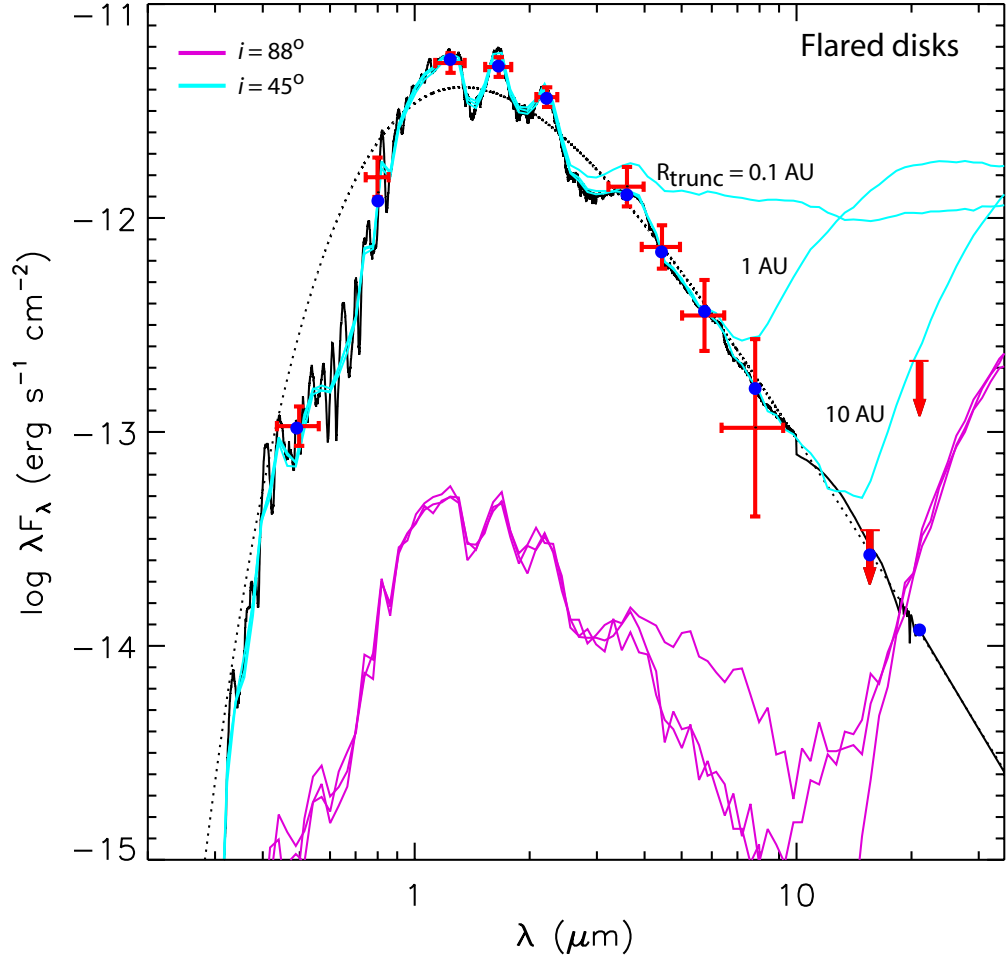}
\caption{\label{flared-sed-fig}
Same as Fig.~\ref{knife-sed-fig}, but with magenta solid curves now 
representing model SEDs of flared disks. Blue curves represent these 
same flared disks seen at a more face-on inclination of $45^\circ$.
The magenta curves have not been scaled differently relative to 
Fig.~\ref{knife-sed-fig}; rather, the photospheric flux is heavily
suppressed due to the heavy obscuration by the edge-on flared disk.
The sampling noise in the magenta curves is also due to the heavy 
obscuration by the edge-on disk, resulting in relatively few Monte Carlo
photons from the photospheres passing unimpeded to the observer.
}
\end{figure}

Fig.~\ref{obs-sed-fig} compares our observed optical-to-MIR photometry
to {\sc dusty} synthetic spectra at the spectral-type-dependent \teff\
adopted by S07 (and supported by R07), namely, $\left[T_1, T_2\right] = \left[2715{\rm
K}, 2890{\rm K}\right]$. The match is clearly very good, with the
predictions agreeing with the data to within $\sim 1 \sigma$ in all bands.

\begin{figure}[ht]
\epsscale{0.8}
\plotone{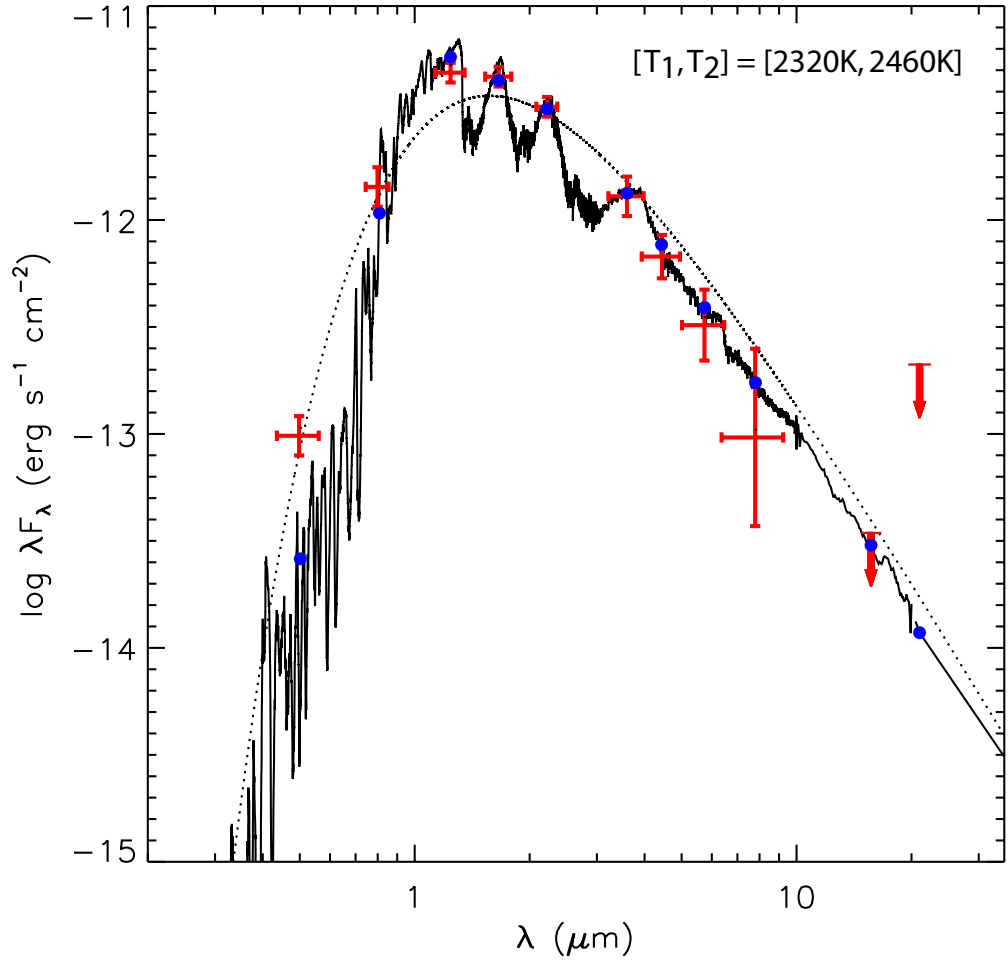}
\caption{\label{cool-sed-fig}
Same as Fig.~\ref{obs-sed-fig}, but with model atmosphere temperatures
of 2320~K and 2460~K for the primary and secondary brown dwarfs,
respectively. To best fit the data, the model has been scaled to a 
distance of 330 pc with no reddening.
}
\end{figure}

Fig.~\ref{cool-sed-fig} shows the analogous comparison at the cooler
\teff\ proposed by CGB07: $\left[T_1, T_2\right] = \left[2320 {\rm
K}, 2440 {\rm K}\right]$. In the NIR and MIR, the agreement with the
synthetic photometry is just as good as in the hotter case above;
the reduced $\chi^2$ of the fit at $1 \le \lambda \le 8 \mu$m is 
$\chi_\nu^2 = 1.06$ with a best-fit distance of $330 \pm 11$ pc and 
$A_V = 0.0 \pm 0.1$. However, even at so near a distance and with no 
extinction, in the $g'$-band there is a large discrepancy, with the
predicted photometry fainter than the data by $\sim 4.5 \sigma$. Basically,
the observed SED appears significantly bluer than expected at the CGB07
temperatures. {\it Prima facie}, these results support the hotter 
\teff\ suggested by S07/R07.
 
One may argue that the synthetic spectra suffer from opacity
uncertainties that considerably weaken the above comparisons and
conclusion. Given the huge number and complexity of atomic and
molecular opacities that must be accurately modeled to reproduce
such cool spectra, this objection is worthy of consideration. A
counter-argument might be that we find the optical data to be
bluer than predicted in Fig.~\ref{cool-sed-fig}, while shortcomings
in the opacities have led, if anything, to the observed optical colors
(e.g., $V-I$) of field M dwarfs being {\it redder} than in the
models \citep[e.g.,][]{Allard00}. As those authors also show, newer
opacities---incorporated in the models we use---have now substantially
mitigated this discrepancy. Nevertheless, it is fruitful to attack
the problem via more general considerations.

We do so by comparing our data to blackbody curves at the suggested
temperatures. M-type spectra clearly depart significantly from
pure blackbodies, but the precise {\it manner} in which they do
so provides important insights. In particular, molecular opacities
(predominantly TiO and VO) strongly suppress the optical spectrum,
relative to a blackbody at the same \teff, in early- to mid-M dwarfs;
dust opacity exacerbates this effect in the late M types. A smaller
suppression of flux also occurs over a large swathe of the MIR due to
H$_2$O absorption. For a given \teff, this flux must escape somewhere,
and it does so predominantly in the NIR, leading to an enhancement of
flux, relative to a blackbody, at these wavelengths. The upshot is
that for M type objects, a blackbody at the same \teff\ represents
an upper limit to the actual emitted flux in the optical and MIR,
and a lower limit in the NIR. This is a general prediction of basic
opacity considerations in these objects, independent of the precise
details of synthetic spectra and attendant opacity uncertainties.

Figs.~\ref{obs-sed-fig} and \ref{cool-sed-fig} show how our data
compare to blackbodies at the S07/R07 and CGB07 \teff\
respectively. In both plots, the blackbodies have been scaled
to the same distance and have been reddened by the same amount as 
the synthetic spectra, i.e., the relative
differences between the two are preserved. At the hotter proposed
\teff\ (Fig.~\ref{obs-sed-fig}), where the synthetic spectra agree
with the observations in all bands, the blackbody follows the
prediction above: in the optical and MIR it is brighter than, and in
the NIR fainter than or at most equal to, the observed (and synthetic)
photometry. At the CGB07 \teff\ (Fig.~\ref{cool-sed-fig}), however, the
situation is very different: while the blackbody is somewhat brighter
than the data in the MIR and fainter than or equal to it in the NIR,
as expected, in the optical $g'$-band the blackbody {\it matches} the
data (which is more than 4$\sigma$ brighter than the synthetic $g'$,
as stated earlier). In other words, if this were the correct \teff,
the synthetic spectra would have to produce flux {\it at blackbody
levels} in the optical $g'$-band to reproduce the data\footnote{One
cannot escape this conclusion by invoking a change in scaling, such
that the synthetic spectrum at the CGB07 \teff\ approaches the optical
data more closely, and the blackbody falls above the latter: any
significant rescaling in this direction will push the blackbody curve
far above the NIR observations as well, which is also inadmissible by
our simple arguments above.}. This is very much against expectations
for M type objects, as outlined above. These results again indicate
that the hotter \teff\ proposed by S07/R07 better represent the true
situation than the significantly cooler ones proposed by CGB07. 
At the very least, we can conservatively state that the CGB07 \teff s are
a lower limit, and reality is likely to be closer to the S07/R07 results. 

It is somewhat more difficult to place an {\it upper} limit on the
component \teff s based on our broad-band SED measurements. At warmer 
\teff s, the model SEDs show bluer colors and produce higher luminosities 
(for fixed radii).  Consequently, to accommodate warmer temperatures,
the system must be placed at a larger distance and must be seen through
larger amounts of extinction. Meanwhile, the model SEDs change relatively
little at $\lambda > 1 \mu$m. Thus it is possible to fit the observed
SED very well with virtually any warmer \teff\ so long as the system
is highly reddened and is placed at a suitably high distance. However, 
if we assume that the system can be no further than 520 pc and reddened 
by at most $A_V = 2.0$, then the maximum temperatures that still allow
the model SED to fit the data within 95\% confidence are 
$\left[T_1,T_2\right] = \left[3230{\rm K},3435{\rm K}\right]$, where
again we adopt the \teff\ ratio and radii from S07. We
emphasize that this upper limit is based solely on the broad-band SED;
such high \teff s would not be consistent with the spectral typing
of SMV06 or the high-resolution spectral modeling of R07, which indicate
maximum \teff s of 
$\left[T_1,T_2\right] \approx \left[2850{\rm K},3000{\rm K}\right]$.

\section{Discussion and Conclusions}
Our {\it Spitzer} observations rule out optically thick circumstellar
disks around the individual components of \eb. While the data cannot
completely rule out a optically thick circumbinary disk, due to the upper
limits in the 16 and 24$\mu$m photometry, any such disk must have a large
inner hole: $\gtrsim 10$ AU inner radius if the disk is flared, or $\gtrsim
0.6$ AU if it is flat. 
These dust hole sizes would imply clearing by either significant grain
growth and evolution, which in turn makes it likely that the disk is near
the end of, or beyond, its main accretion phase; or by planetesimal
formation or photoevaporation, both of which would clear gas as well as
dust within the hole, also making significant ongoing accretion unlikely.
We also stress that the derived inner-disk hole sizes are {\it lower limits}
based on our 16 and 24$\mu$m photometry upper limits; thus it is very likely
that the holes are significantly bigger (or indeed that there is no disk at all). 
Accretion is also unlikely given that
(a) it is the primary that shows large H$\alpha$ emission (while binary
accretion theory predicts preferential accretion onto the secondary),
and (b) there is no evidence for the type of large-amplitude photometric
variability usually observed in accreting systems \citep{gomez08}. 
In combination, these results indicate that it is not accretion but
chromopsheric activity that underlies the strong $\hal$ emission from the
\eb\ primary. In turn, this bolsters the argument, made by CGB07 and R07,
that the \teff\ reversal between the components is due to the combined
effects of the rotation and magnetic fields that underlie activity. We
now discuss the S07/R07 and CGB07 \teff\ results in this light.

\subsection{Problems with the cooler effective temperatures predicted 
for \eb}

There are two important reasons for trying to distinguish between the
hotter \teff s for \eb\ determined by S07/R07 and the much cooler
ones proffered by CGB07. The first concerns activity. CGB07 are
able to match SMV06's mass-radius data by invoking, in the primary,
both a large global reduction in convective efficiency (mixing length
parameter $\alpha = 0.5$ instead of the usual $\approx 2$) and a large
covering fraction of cool spots $\beta$ = 0.5, leading to a primary
\teff\ $= 2320$K. At least in field dwarfs, however, such \teff\ are
typical of $\gtrsim$ M9 dwarfs, which are usually highly {\it in}active despite
rapid rotation \citep[e.g.,][]{mohanty03}. The generally adopted reason
for this is a rapid drop in the photospheric ionization levels at
such low \teff\ \citep{mohanty02}. This phenomenon appears to extend
to M dwarfs in star-forming regions as well, where the same reduction
in activity (e.g., in X-rays) is seen with later M-types, and ascribed
to the same physics \citep[e.g.,][]{stelzer05}.

Adopting the CGB07 \teff\ would thus imply that the \eb\ primary is
somehow able to remain extremely active despite being very cool. CGB07
acknowledge this problem, but their solution---activity started on
the primary when it was much hotter, but the resulting suppression of
convection eventually drove its \teff\ down to its present value---does
not appear very satisfactory: the primary still has to {\it maintain}
very strong activity at its {\it current} \teff\ to explain the
(observed) large $\hal$ emission and (proposed) very high spot covering
fraction. The S07/R07 \teff, corresponding to usual mid-M values,
would be much more consistent with such activity / spot coverage:
activity levels peak in the mid-M types \citep[e.g.,][]{gizis02}.


The second issue with the lower \teff s suggested by CGB07 concerns the
fundamental parameters assigned to brown dwarfs and low mass stars in
general, based on their \teff\ and luminosity. To illustrate this, we plot
both \eb\ components on the HR diagram, at the positions corresponding to the
S07 and CGB07 \teff\ estimates (Fig.~\ref{subu-fig}). The theoretical
evolutionary tracks and isochrones plotted are from \citet{baraffe98}
and \citet{Chabrier00}. For the hotter S07 \teff\ values, it is
evident that the primary appears less massive than it really is (0.054
M$_{\odot}$) and also much younger than expected ($\sim$ 1 Myr), while
the secondary appears more massive than its true value and slightly older
than expected. The disparity in the secondary's position can be explained
if the evolutionary tracks are slightly ($\sim 100$ K) too cold, or the
spectral type to \teff\ conversion is a little too hot (or, probably,
a combination of both). Note that such a correction would simultaneously {\it
enhance} the disparity in the primary's position (since the \teff\ ratio of
the secondary to primary is empirically fixed): this is of course precisely
the empirically observed reversal in \teff\ we are seeking to explain.
At any rate, the point is that the secondary's position in the HR diagram
is not very far removed from its real one if one uses the \teff\ suggested
by its spectral type.

CGB07's \teff, however, imply a much more extreme picture: not only is the
primary far removed from its true mass/age, but {\it so is the secondary}
(i.e., the disparity in the secondary's position is much larger than in the
above case, where its \teff is derived from spectral type).  In other words,
in this model {\it even slowly rotating, relatively inactive young brown
dwarfs are very sensitive to rotation/B-field effects}.  (CGB07 adopt only
a moderate global suppression of convection in the secondary: $\alpha =
1$, and a low cool spot coverage: $\beta = 0.2$, in keeping with its low
\vsini and weak $\hal$ emission.)  Note that this conclusion cannot be
avoided by appealing to uncertainties in spectral type / \teff\ conversion or
disparities between evolutionary tracks by different groups. The theoretical
tracks plotted are the same ones constructed by a subset of the CGB07 authors
in the absence of rotation/B-field effects, and the same ones used by CGB07
as a baseline to include the latter physical effects. Fig.~\ref{subu-fig}
shows that the \teff\ {\it predicted by these tracks}, for the known mass
and assumed age of the secondary, is {\it (a)} close (within $\sim$150K)
of the \teff\ S07 find from spectral type considerations, and {\it
(b)} much higher (by $\sim$300K) than the secondary \teff\ derived by
CGB07 by including its minor rotation/B-field. The inescapable conclusion
is that even a small rotation/field has a very large effect on evolution
in CGB07's treatment.


\subsection{Implications for theoretical models of young low-mass
stars and brown dwarfs}

All our above analyses---both from the observational standpoint of
SEDs, and the theoretical ones of plausibility of strong activity
on the primary and agreement of the slowly rotating, weakly active
secondary with the isochrones---suggest that the hotter \teff\
proposed by S07/R07 are more likely for \eb's components than the
cooler ones predicted by CGB07 by including rotation/field effects.  Under the circumstances, we briefly examine the viability of two other potential mechanisms for producing the temperature reversal.  

{\it Accretion}:
Disk accretion can change the properties (radius, luminosity)
of T Tauri stars; while the components of \eb\ appear to be post-accretion, is 
it possible that the effects of their past accretion phase (which must have occurred) linger, and give rise to the observed temperature reversal?
The brief answer is that this is quite improbable, for the following
reason.  In most T Tauri stars, with accretion rates $\lesssim 10^{-5}\;
{\rm M}_\odot\; {\rm yr}^{-1}$, the accretion is assumed to be ``cold''.
That is, the thermal energy of the accreted material is radiated away into space from the uppermost photospheric layers over a small fraction of the stellar
surface, and contributes negligibly to the internal energy of the star
\citep{hartmann97}.  It is only in the extreme conditions and very
high accretion rates found in FU Orionis-type objects that this condition
is likely to be violated.  In the ``cold'' limit, the effect of accretion
(when the accretion timescale becomes comparable to, or shorter than, the
Kelvin-Helmholtz timescale) is to decrease the stellar luminosity by
decreasing the {\it radius}, without affecting the stellar temperature
very much: essentially, the star does not have time to expand to the
radius it would have for the corresponding mass in the absence of
accretion.

In the case of the BD components under discussion here, the past accretion
for each should have been of order \mdot $\lesssim 5\times 10^{-7}$
\msun yr$^{-1}$ (given their masses of $\sim$ 0.05 \msun, and assuming
a minimum age of 10$^5$ years).  Inserting this along with their other
properties into the equations supplied by \citet{hartmann97} shows
that the ``cold'' limit is applicable to our BDs as well, as might have
been intuitively expected from the much smaller accretion rates in BDs compared
to T Tauri stars.  One does not therefore expect the accretion to change
their temperatures by any significant amount, only their radii.  Recent
calculations assuming ``cold'' accretion in BDs bear this out: accretion
rates of $\gtrsim 10^{-8}$ \msun yr$^{-1}$ can decrease the BD radius
significantly (up to few tens of percent), thus making it sub-luminous
compared to a non-accretor of the same mass, but the temperature of the BD
remains virtually unchanged from its value in the absence of accretion
\citep[][see also discussion of the latter results in 
Chabrier et al.\ 2007]{gallardo08}.  
Therefore, while future calculations should further clarify the effects of accretion on BD properties, it appears quite unlikely that the temperature reversal observed in \eb\ is caused by the lingering effects of past accretion.

{\it Non-coevality}:
Second is the intriguing possibility that the assumption of coevality for the binary components may not be quite true. From comparisons to theoretical tracks, S07 concluded that an age difference of $\sim$ 0.5 Myr, together with a slight ($\sim$ 70K) reduction in the \teff\ derived from spectral types, could in fact explain the entirety of the \teff\ reversal without resorting to any other mechanism.  

On the one hand, the reality of non-coevality effects in determining initial binary properties is strongly suggested by another eclipsing binary, also in the ONC, discovered recently by \citet{stassun08}.  The stellar components of this system have almost exactly the same mass (0.41 M$_{\odot}$) but differ greatly in \teff\ (by $\sim$300K) and luminosity (by a factor of $\sim 1.5$). With both stars exhibiting very small H$\alpha$ emission, indicating very low field strengths, as well as nearly the same \vsini, field and rotation effects do not appear plausible for explaining their differences. However, a difference of a few hundred thousand years in the individual ages of the stars can indeed do the trick, as Stassun et al.\ postulate.  A similar non-coevality in \eb\ therefore cannot be discounted.

At the same time, the reality of field/rotation effects on low mass objects is also strongly supported by recent data.  In particular, \citet{morales07} show that very active field dwarfs are cooler and larger than their non-active counterparts at the same luminosity (or equivalently same mass, given the tight mass-luminosity correlation on the low-mass ZAMS).  Given that the components of \eb\ have the same mid-M spectral type and interior structure (fully convective) as the low-mass end of Morales et al.'s sample, do exhibit strong differences in rotation and activity, and evince precisely the same trend (more active component is less luminous and larger than expected) as the field dwarfs, the role of activity/magnetic fields in determining the properties of \eb\ cannot be discounted either.  

It therefore appears plausible that both age and rotation/field effects are important for understanding the \teff\ reversal in \eb.  Indeed, this may resolve the problem of uncomfortably low \teff\ derived in the CBG07 analysis: if non-coevality (of $<$ 0.5 Myr) moves the system {\it partially} towards \teff\ reversal, the remainder of the reversal may be effected by rotation/field effects using temperatures more in line with the higher values espoused here and by S07/R07.           

\begin{figure}[ht]
\epsscale{1.00}
\plotone{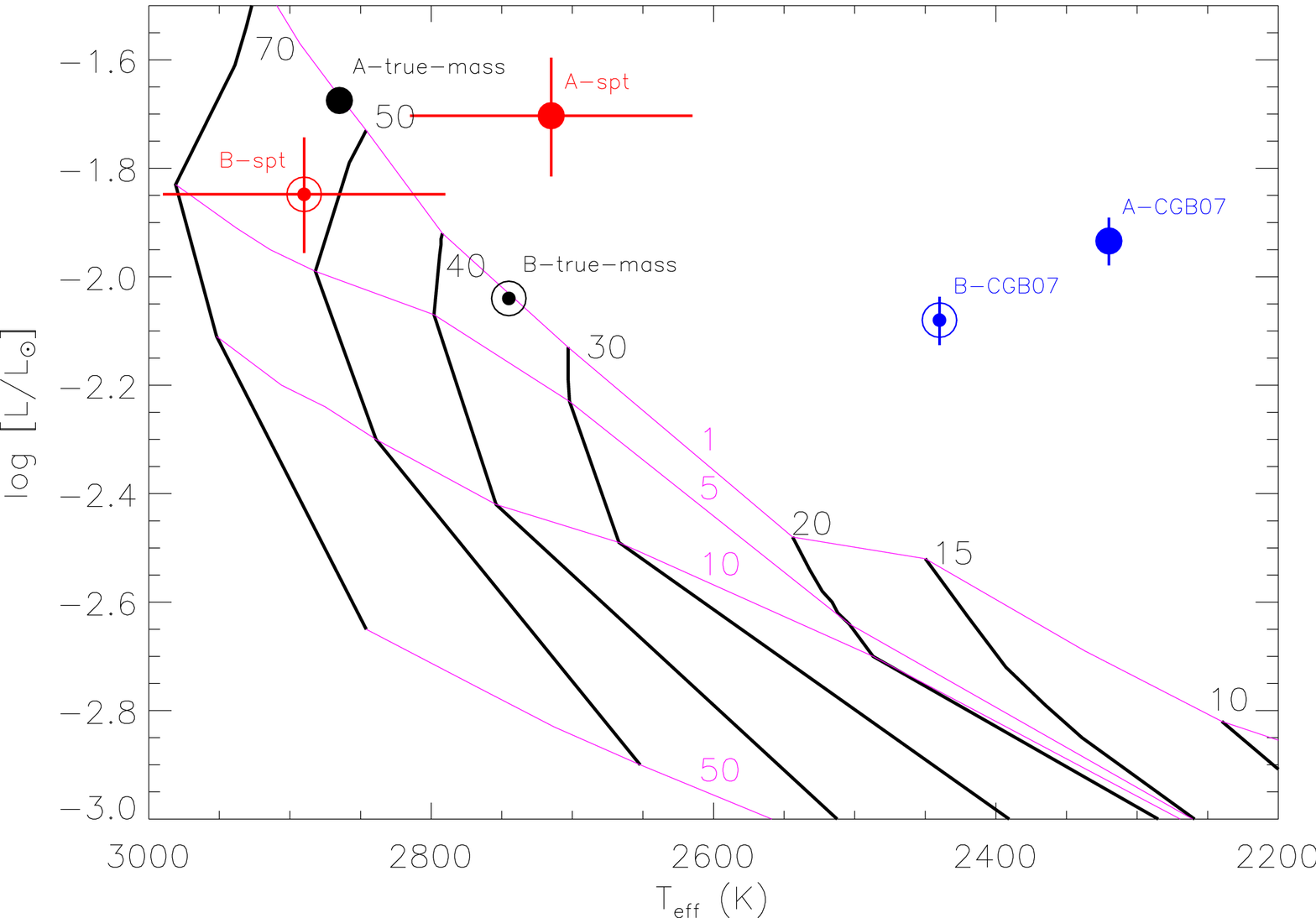}
\caption{\label{subu-fig} 
The observed \teff\ and luminosities of the components of \eb\
are compared to theoretical predictions on a Hertzsprung--Russell
diagram. Black curves are mass tracks at the (Jovian) masses
indicated, magenta curves are isochrones at the ages indicated in
Myr \citep{baraffe98,Chabrier00}. Symbols represent the \eb\ primary
(filled) and secondary (bulls-eye). Black symbols represent the predicted
values for coeval brown dwarfs of the measured masses at 1 Myr. Red
symbols represent the observed values using the \teff\ determined by
S07 and R07. Blue symbols represent the
\teff\ predicted by the models of \citet{chabrier07}.
} 
\end{figure}

Finally, we emphasize the importance of pinning down the potential effects of rotation/fields in young low-mass objects for determining the overall IMF of low-mass stars and brown dwarfs.  A large fraction of young, non-accreting brown dwarfs rotate significantly faster than not only the \eb\ secondary, but than the primary as well; many of these also have H$\alpha$ EWs, and thus presumably field strengths, comparable to the primary's \citep{Mohanty05}. If rotation/field effects are important, as the results for \eb\ and the field dwarfs suggest, the masses of these objects obtained from comparisons to the usual evolutionary tracks (which do not include such effects) are possibly significantly incorrect (analogous to the situation for the \eb\ primary; see Fig.~\ref{subu-fig} and \S5.1), and thus so is the derived IMF. \eb\ is an extremely  fortuitous find in this regard, not only because it is an EB, but also because its two components represent both classes of brown dwarfs: rapidly rotating highly active ones, and slowly rotating weakly active ones. A good physical understanding of both components is essential for an accurate description of the fundamental properties of brown dwarfs and low-mass stars.

\acknowledgments
S.M.\ is grateful to the Spitzer Fellowship program for funding this
research. K.G.S.\ acknowledges funding support from a Spitzer Cycle-4 GO
grant and a Cottrell Scholar award from the Research Corporation.

\clearpage

\clearpage

%

\clearpage

\clearpage

\clearpage

\clearpage

\begin{deluxetable}{lcc}
\tablecaption{Broadband Flux Measurements of \eb\label{obs-table}}
\tablecolumns{3}
\tablewidth{0pt}
\tablehead{
\colhead{Passband} & 
\colhead{$\lambda$} & 
\colhead{$F_\lambda$} \\
\colhead{} & \colhead{$\mu$m} & \colhead{mJy} 
}
\startdata
$g'$   &   0.48   &   0.017$\pm$0.002 \\
$I_C$  &   0.79   &   0.35$\pm$0.03 \\
$J$    &   1.24   &   2.20$\pm$0.10 \\
$H$    &   1.66   &   2.82$\pm$0.13 \\
$K_S$  &   2.22   &   2.73$\pm$0.13 \\
IRAC1  &   3.6    &   1.69$\pm$0.16 \\
IRAC2  &   4.5    &   1.10$\pm$0.11 \\
IRAC3  &   5.8    &   0.68$\pm$0.11 \\
IRAC4  &   8.0    &   0.28$\pm$0.12 \\
IRS PU &    16    &   $<0.18$\tablenotemark{a} \\
MIPS   &    24    &   $<1.69$\tablenotemark{a} \\
\enddata
\tablenotetext{a}{3$\sigma$ upper limit.}
\end{deluxetable}

%
%

%
%
%
%
%
%
%

\end{document}